\documentclass[preprint,12pt]{elsarticle}
\usepackage{amssymb}
\usepackage{amsmath}
\usepackage{xcolor}

\usepackage{booktabs}
\usepackage{hyperref} 
\usepackage{tabularx}
\usepackage{array}
\usepackage{graphicx}
 \usepackage{longtable} 
\usepackage{subcaption}
\usepackage[gen]{eurosym}
\usepackage{xcolor}  

\usepackage{url}
\usepackage{lscape}
\usepackage{pdfpages}
\makeatletter
\def\@biblabel#1{\hbox{#1.}}
\makeatother

\begin{document}
\begin{frontmatter}

\title{Causal discovery identifies pathways linking physical activity to dementia risk  in the UK BioBank}

\author[1]{Wasif Khan}
\author[2]{Panayiotis V. Benos}

\author[5]{Joshua K. Wong}

\author[1,7,9]{Ruogu Fang\corref{Ruogu Fang}}

\affiliation[1]{
    organization={J. Crayton Pruitt Family Department of Biomedical Engineering, University of Florida},
    city={Gainesville},
    state={FL},
    country={USA}
}

\affiliation[2]{
    organization={Department of Epidemiology, University of Florida},
  city={Gainesville},
   state={FL},
    country={USA}
}



\affiliation[5]{
    organization={Norman Fixel Institute for Neurological Diseases, University of Florida},
    city={Gainesville},
    state={FL},
    country={USA}
}

\affiliation[7]{
    organization={Center for Cognitive Aging and Memory, McKnight Brain Institute, University of Florida},
    city={Gainesville},
    state={FL},
    country={USA}
}
\affiliation[9]{
    organization={Department of Electrical and Computer Engineering, University of Florida},
    city={Gainesville},
    state={FL},
    country={USA}
}

\cortext[Ruogu Fang]{Corresponding author}
\ead{ruogu.fang@bme.ufl.edu}

\begin{abstract}
Physical Activity (PA) is consistently associated with lower risk of dementia, yet the mechanism linking PA to dementia prevention remain incomopletely understood. Here, we integrate large language model (LLM)-guided causal discovery with mediation analysis in 42,293 older adults aged 60 years or older from the UK Biobank to systematically identify pathways connecting objectively measured moderate-to-vigorous physical activity (MVPA) to dementia risk. Across behavioral, psychological, functional, and clinical domains, causal discovery consistently identified interconnected pathways linking higher MVPA to lower dementia risk through depression, functional capacity, smoking behavior, hypertension, cardiovascular disease, chronic kidney disease, and brain injury. Chain mediation analyses further identified depression as a central pathway, accounting for 15.1\% of the overall association between MVPA and dementia risk. Sex-stratified analyses revealed distinct mechanistic patterns, with females exhibiting predominantly metabolic and functional pathways, while males showed behavioral and cardiometabolic cascades involving smoking and cardiovascular disease. Together, these findings suggest that the protective association between PA and dementia is mediated through interconnected behavioral, psychological, and cardiometabolic processes rather than a single pathway. Depression emerged as a prominent and potentially modifiable pathway, highlighting opportunities for integrating dementia prevention strategies that combine PA promotion with mental health and cardiovascular risk management.
\end{abstract}

\end{frontmatter}

\section{Introduction}

Dementia is a major global public health challenge, affecting more than 57 million people worldwide and accounting for nearly 10 million new cases each year \cite{WHO_Dementia_2025}. As a leading cause of disability, dependency, and mortality among older adults, dementia imposes an annual global economic burden exceeding US \$1.3 trillion. In the absence of disease-modifying therapies, identifying modifiable factors that delay or prevent dementia onset has become a critical public health priority.

Among the modifiable risk factors for dementia, physical activity (PA) is one of the most consistently identified protective lifestyle behaviors. The 2024 Lancet Commission estimated that approximately 45\% of dementia cases could be prevented by addressing 14 modifiable risk factors \cite{khan2025association,livingston2020dementia}. Large prospective cohort studies and meta-analyses have consistently shown that higher levels of PA are associated with reduced risks of cognitive decline and dementia \cite{wanigatunga2025moderate,buchman2012total,petermann2021dose,wu2023dose}. For example, a meta-analysis of 44 prospective cohort studies reported that engaging in PA for at least one hour twice weekly was associated with a 23\% lower risk of all-cause dementia and a 32\% lower risk of Alzheimer's disease \cite{park2023modifying,lee2018relationship}. More recently, accelerometer-based analyses demonstrated that even relatively low levels of moderate-to-vigorous physical activity (MVPA) were associated with nearly a 50\% lower risk of all-cause dementia compared with complete inactivity \cite{wanigatunga2025moderate}. Intervention studies further suggest that multidomain PA programs improve cognitive outcomes, with potentially stronger benefits among women \cite{barha2017sex}. Together, these findings establish PA as a promising target for dementia prevention.

Despite compelling epidemiological evidence, the mechanisms through which PA reduces dementia risk remain incompletely understood. Multiple biological and behavioral pathways have been proposed. At the biological level, PA has been associated with improved cerebrovascular function and cerebral perfusion, reduced systemic inflammation, enhanced cardiometabolic health, and increased neuroplasticity, all of which may contribute to resilience against neurodegeneration \cite{klil2026physical, guadagni2020aerobic, ben2025physical, bruunsgaard2005physical, warnberg2010physical,thorand2006sex,ekelund2012moderate,henson2013associations,blodgett2024device}. PA also influences behavioral and psychosocial factors linked to dementia risk, including depression, mobility, frailty, sleep quality, and social engagement \cite{khazaie2023effect,su2021role,ghrouz2019physical,lampinen2003reduced,pahor2014effect,peluso2005physical,wankel1990psychological}. Rather than acting independently, these biological, psychological, and functional processes likely interact as an interconnected network. Consequently, whether PA exerts direct neuroprotective effects or primarily influences dementia risk through intermediate pathways remains unclear.

Addressing this question requires analytical approaches that move beyond conventional association-based analyses. Although regression and mediation models have identified individual mediators, they generally require predefined hypotheses and are limited in their ability to reconstruct complex dependency structures among numerous interacting risk factors. Causal discovery methods offer a data-driven framework for inferring potential causal relationships from observational data, but their application to large-scale biomedical datasets remains challenging because they rely heavily on manual variable selection and do not leverage the semantic relationships encoded in heterogeneous clinical variables. Recent advances in large language models (LLMs) provide an opportunity to overcome these limitations by incorporating domain knowledge from variable descriptions to facilitate biologically informed feature selection and causal inference.

Here, we integrate LLM-guided feature selection, causal discovery, and mediation analyses to systematically investigate pathways linking objectively measured MVPA to dementia risk in 42,293 older adults from the UK Biobank. Using behavioral, psychological, functional, and clinical variables, we reconstruct interconnected pathways through which PA may influence dementia risk, quantify pathway-specific contributions, and evaluate the robustness of inferred relationships through bootstrap and sensitivity analyses. By moving beyond conventional association-based analyses, our study provides a systems-level understanding of the mechanisms linking physical activity to cognitive aging and identifies potentially modifiable intervention targets for dementia prevention.

\begin{figure}
    \centering
    \includegraphics[width=\linewidth, trim={0cm 5cm 1cm 1cm}, clip]{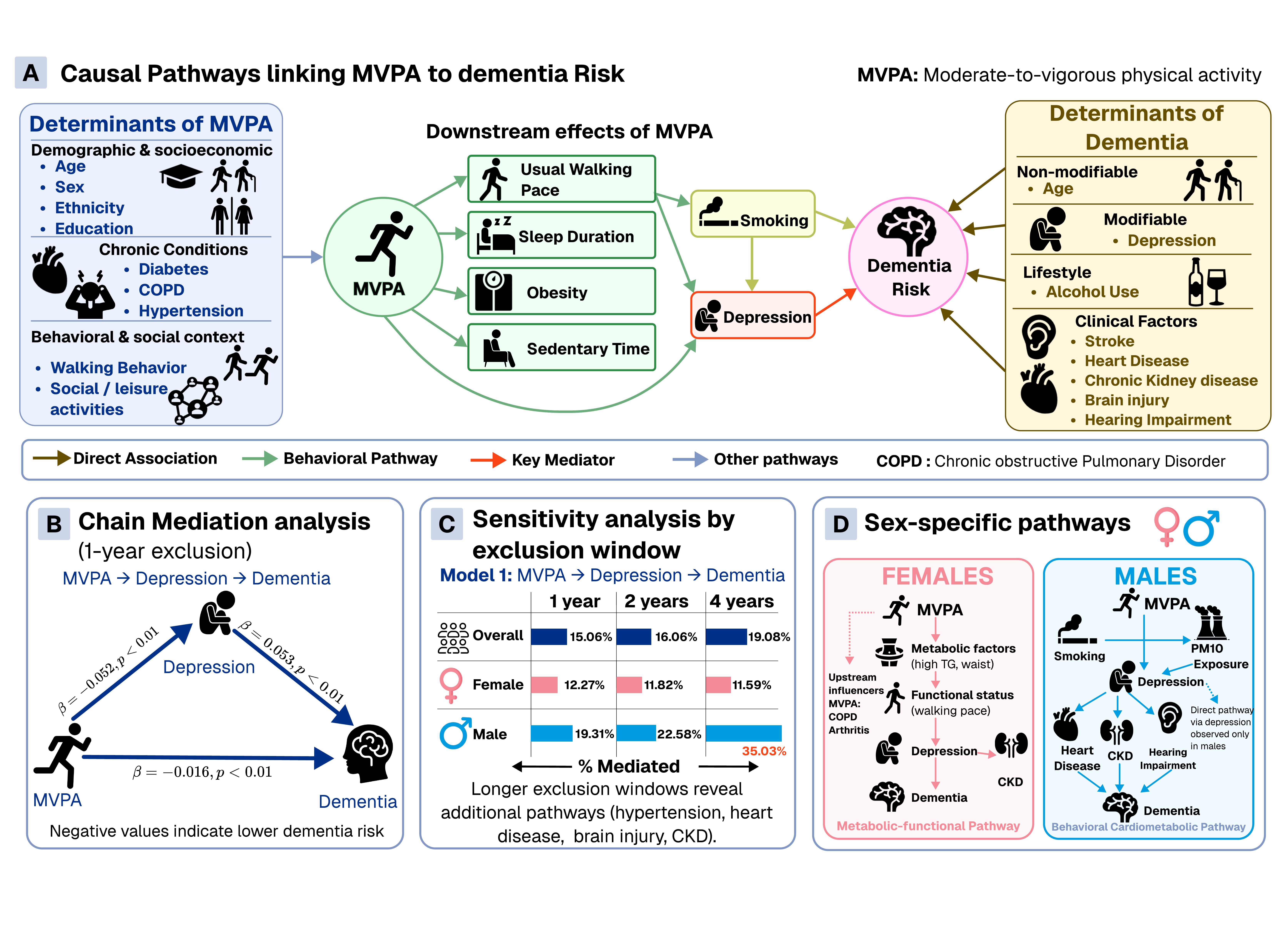}
    \caption{Causal pathways linking moderate-to-vigorous physical activity (MVPA) to dementia risk. (A). Causal graph derived from the PC causal discovery algorithm illustrating inferred pathways linking MVPA to dementia risk in older adults from the UK Biobank. Demographic, socioeconomic, behavioral, and clinical variables were incorporated into the causal discovery framework under domain-informed structural constraints. The resulting network identified both direct and indirect pathways connecting MVPA with dementia risk, including behavioral, functional, and psychological intermediates. Highlighted pathways indicate key mediation chains involving walking pace and depression. (B). Chain mediation analyses quantifying direct and indirect effects of MVPA on dementia risk through depression pathway.  (C). Sensitivity analyses examining mediation effects across progressively longer exclusion windows (1-, 2-, 4-, and 5-year exclusions) to minimize potential reverse causation. Longer exclusion periods identified additional pathways involving hypertension, heart disease, brain injury, and chronic kidney disease (CKD). (D). Sex-specific causal pathways linking MVPA to dementia risk. Female-specific pathways primarily involved metabolic-functional mechanisms, including triglycerides, waist circumference, walking pace, depression, heart disease, and CKD, whereas male-specific pathways additionally involved smoking, PM10 exposure, hearing impairment, and a direct depression-mediated pathway. COPD, chronic obstructive pulmonary disease; CKD, chronic kidney disease; TG, triglycerides.}
    \label{fig:Fig1}
\end{figure}

\section{Results}
Baseline characteristics of the 42,293 participants from the UKB aged $\geq60$ years with valid accelerometer-derived MVPA data are presented in Table 1. We identified a final set of 39 variables for downstream causal and mediation analyses following a multi-stage LLM-guided selection pipeline and expert curation (Supplementary Table 1). The mean age of participants was 67.0±4.1 years, and 53.3\% were female. Overall, participants accumulated an average of 40.2±34.0 minutes/day of MVPA with males demonstrated higher MVPA (48.0 vs 33.3 min/day), and higher prevalence of health conditions, including heart disease, hypertension, diabetes, stroke, chronic kidney disease, and hearing impairment. Incident dementia occurred in 647 participants (1.5\%) after one year of the accelerometer study, with a higher incidence observed in males than in females (1.9\% vs. 1.2\%).

\subsection*{Causal pathways linking physical activity to dementia risk}
Causal discovery analyses identified robust, directionally consistent pathways linking MVPA to dementia risk through behavioral, functional, and psychological processes (Supplementary Fig. 1, Data 1 Table A1). Across bootstrap-resampled models, a primary mediation pathway was observed in which MVPA influenced dementia risk via depression with high structural stability (Fig. \ref{fig:Fig1}, Supplementary Fig. 1, Data 1 Table A2). Furthermore, MVPA was positively associated with usual walking pace, which in turn influenced depression and subsequent dementia risk (Fig.~\ref{fig:Fig1}). An extended behavioral cascade was also observed in which MVPA was linked to walking pace, which was associated with smoking behavior, subsequently influencing depression and dementia risk. Moreover, MVPA was primarily influenced by demographic and socioeconomic factors, including age, sex, ethnicity, and education, and was lower among individuals with chronic conditions such as diabetes, chronic obstructive pulmonary disease, and hypertension. In contrast, higher MVPA was associated with favorable health characteristics, including engagement in social and leisure activities, lower adiposity, better sleep patterns, and faster walking pace, suggesting potential behavioral, functional, and psychological pathways relevant to dementia risk. Within the dementia-related subgraph, age was the dominant determinant of risk, whereas depression emerged as modifiable factor directly connected to both MVPA and dementia. This suggest that depression may represent a central pathway through which PA influences dementia risk and thus provided the rationale for subsequent mediation analyses.Additional contributors included alcohol use and major comorbidities such as stroke, heart disease, chronic kidney disease, brain injury, and hearing impairment.


\begin{figure}
    \centering
    \includegraphics[width=\linewidth, trim={0cm 3cm 1cm 1cm}, clip]{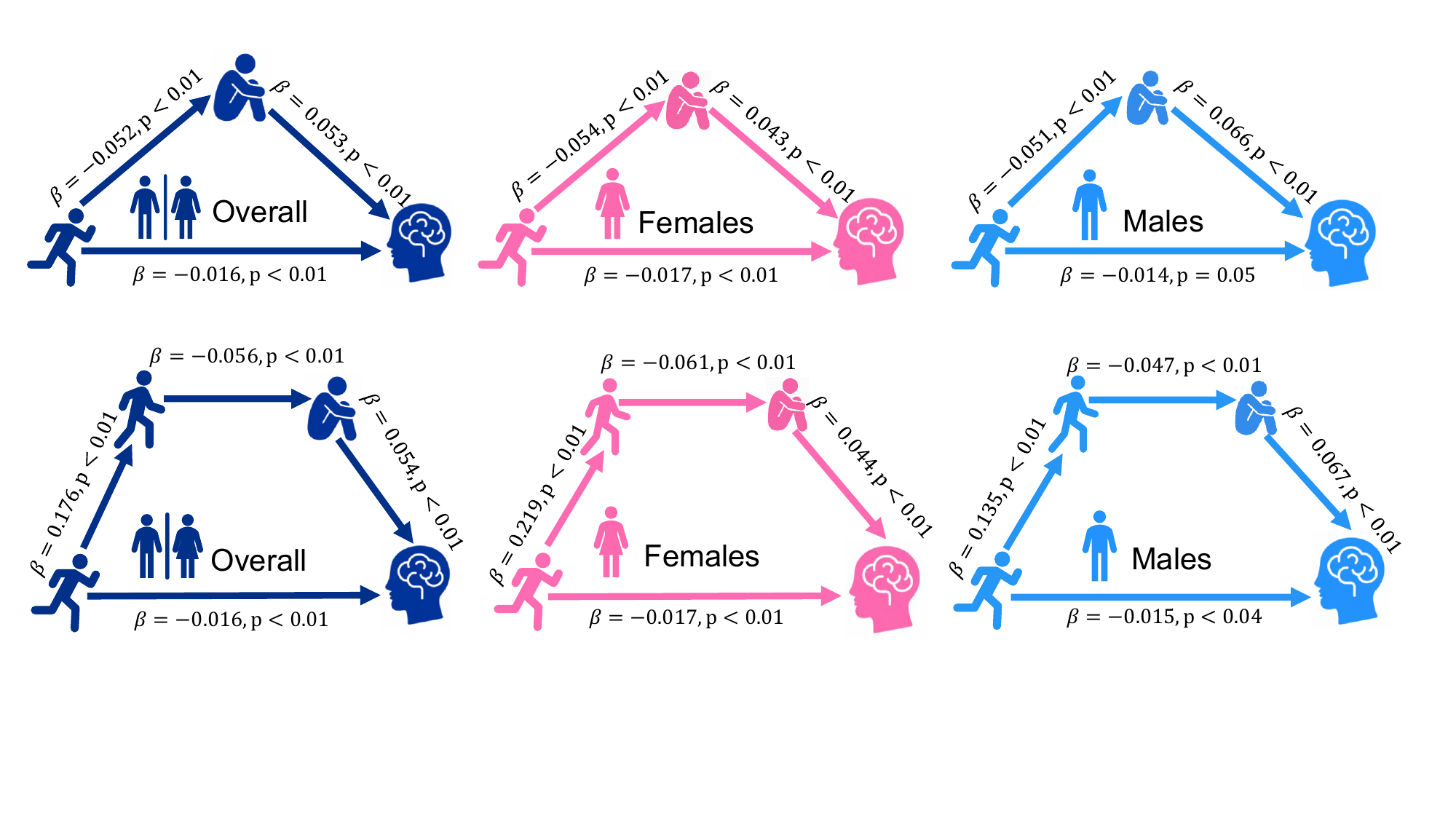}
    \caption{Chain mediation analyses examining the indirect pathways linking moderate-to-vigorous physical activity (MVPA) with dementia risk through depression and walking pace. The left panels represent the overall population, the middle panels represent females, and the right panels represent males. Upper panels illustrate the mediation pathway of MVPA $\rightarrow$ depression $\rightarrow$ dementia in the overall cohort and stratified by sex (females and males). Lower panels present the extended chain mediation pathway incorporating walking pace, modeled as MVPA $\rightarrow$ walking pace $\rightarrow$ depression $\rightarrow$ dementia. Standardized regression coefficients ($\beta$) and corresponding p-values are shown for each pathway. Models were adjusted for age, sex, ethnicity, education, deprivation index, smoking, and alcohol use.}  
    \label{fig:mediation}
\end{figure}

\subsection*{Chain Mediation Analysis}
Chain mediation analysis quantified direct and indirect pathways of MVPA to dementia risk which revealed that depression partially mediated the association between MVPA and dementia (Fig.~\ref{fig:mediation}, Supplementary Table 2), accounting for 15.06\% ($\beta_{\mathrm{indirect}} = -0.0028$) of the total effect ($\beta_{\mathrm{total}} = -0.0183$). 
Sex-stratified analyses showed a higher proportion mediated in males (19.31\%) compared to females (12.27\%).
Sensitivity analyses demonstrated a progressive increase in the proportion of the MVPA–dementia association mediated by depression with longer exclusion windows (Supplementary Table. S.3). In the overall cohort, the proportion mediated increased from 15.1\% at 1 year to 23.8\% at 5 years with a more pronounced increase in males from 19.3\% at 1 year to 60.8\% at 5 years, whereas estimates in females remained relatively stable (11.6–13.0\%). Furthermore, the modeled direct association between MVPA and dementia progressively attenuated with longer exclusion periods, and was no longer statistically significant after 5 years in the overall cohort and after 2 years in males. In contrast, both the MVPA$\rightarrow$Depression and Depression$\rightarrow$Dementia pathways remained significant across all exclusion periods, consequently, the relative contribution of the depression-mediated pathway increased over time, particularly among males. 
These findings suggest that the depression-mediated pathway remains stable with longer exclusion windows and accounts for an increasing proportion of the observed MVPA–dementia association.
We further evaluated an extended pathway incorporating functional status (MVPA $\rightarrow$ walking pace $\rightarrow$ depression $\rightarrow$ dementia). In this model (at 1 year exclusion of dementia cases), the indirect effect was smaller, accounting for 3.13\% of the total effect, and remained consistent across sexes (females: 3.34\%; males: 2.83\%) and exclusion windows (range: 3--5\%) (Fig.~\ref{fig:mediation}, Supplementary Table 3, Supplementary Fig. S.4). These findings indicate that depression was consistently supported mediator among the pathways examined, whereas the additional indirect pathway incorporating walking pace contributed modestly to the overall association.

\section{Sensitivity and Subgroup Analysis}
Incident dementia cases occurring within 1, 2, 4, and 5 years after accelerometer assessment were excluded, revealing temporal variation in the inferred causal structure (Data 1 Table A1-A8). Depression remained a consistent mediator across all exclusion windows, indicating a stable proximal pathway. Longer exclusion windows (2 and 4 years) revealed additional pathways involving cardiometabolic and neurological intermediates, including hypertension, heart disease, brain injury and chronic kidney disease. These pathways comprised multi-step sequences (Fig. \ref{fig:Fig1}) that were not observed with shorter exclusion windows. Their emergence with increasing temporal separation is consistent with reduced influence of preclinical or early disease processes and greater capture of longer-term pathways. Overall, longer exclusion windows identified additional downstream pathways involving cardiometabolic and neurological conditions linking MVPA to dementia risk. However, these pathways exhibited low bootstrap stability (0.07–0.10), suggesting limited robustness (Supplementary Fig. S.2.)
Sex-stratified causal discovery indicated differences in the network structure relating MVPA to dementia risk, depression was consistently identified as a mediator across sexes. 
In females, MVPA was primarily associated with metabolic and functional variables, including adiposity measures (triglycerides and waist circumference) and functional capacity. In males, the inferred structure comprising multiple pathways directly through depression and via extended sequences involving cardiovascular and clinical conditions, including heart disease, chronic kidney disease, arthritis and hearing impairment. Additional paths involved behavioral and environmental variables, including smoking and particulate matter exposure, as well as sequences through waist circumference and usual walking pace before depression.

\section{Discussion}
Causal discovery and mediation analyses from the UK Biobank suggest that the association between MVPA and dementia risk may operate through multiple behavioral, psychological, and cardiometabolic pathways. Among the pathways examined, depression emerged a consistent pathway across sensitivity analyses and remained evident after excluding incident dementia cases, suggesting that psychological health may represent a potentially modifiable mechanism linking MVPA to long-term dementia risk.
Our findings suggest that part of the protective association of MVPA may be mediated through improvements in mental health. Our results are biologically plausible and in agreement with literature which show that MVPA is associated with reduced stress \cite{ andersen2026curvilinear}, systemic inflammation \cite{ bricca2025device}, and improves neuroendocrine function, which may contribute to depression, neurodegeneration, and dementia risk \cite{ phillips2018immune, su2025effects, hurley2013neuroinflammation, menard2017immune}. Furthermore, pathways incorporating functional capacity (e.g., walking pace) and behavioral factors (e.g., smoking) suggest that MVPA also operates through behavioral pathways, where improvements in functional capacity and lifestyle behaviors influence mental health and cognitive outcomes.  Extended pathways involving cardiometabolic and neurological conditions supports the hypothesis that long-term vascular and systemic health may represent an important link between PA and dementia \cite{dove2024high,wang2020leisure, dove2023cardiometabolic,kivimaki2019physical, kantawala2023physical}. Beyond PA, dementia risk was directly influenced by multiple established factors \cite{livingston2020dementia}, including age, depression, alcohol use, cardiovascular disease, stroke, chronic kidney disease, brain injury, and hearing impairment.

Sensitivity analyses excluding incident dementia cases within 1-5 years demonstrated that the core depression-mediated pathway remain stable, supporting its robustness. Longer exclusion windows revealed additional pathways involving hypertension, heart disease, brain injury, and chronic kidney disease, suggesting cumulative biological processes and long-term mechanisms of neurodegeneration. These findings imply that MVPA may exert both short-term effects through psychological and behavioral pathways and longer-term effects through cardiometabolic and neurovascular mechanisms. Sex-stratified analyses further highlight heterogeneity in these processes. The predominance of metabolic and functional pathways in females may reflect sex-specific differences in body composition, hormonal regulation, and energy metabolism, whereas the more complex behavioral and cardiometabolic pathways observed in males may be influenced by differences in lifestyle patterns and disease burden. However, the consistent role of depression across both sexes suggests a shared psychological pathway that may be broadly relevant for intervention.

Chain mediation analysis quantified the relative contribution of specific pathways linking MVPA to dementia risk, with consistent evidence of both direct and indirect effects. Depression accounted for approximately 15\% of the total effect in the full sample and showing stronger mediation in males ($\sim$ 19\%) compared to females ($\sim$12\%). Chain mediation through cardiovascular disease and functional capacity (walking pace) contributed to the overall association, approximately 4\% of the effect. Sensitivity analyses demonstrated that the proportion of the MVPA–dementia association mediated by depression increased with longer exclusion windows, rising from 15\% at 1 year to 23\% at 5 years. This pattern was more pronounced in males, where mediation increased from 19\% to 60\%, while remaining stable in females ($\sim$ 12\%). 
However, these estimate should be interpreted cautiously because the proportion mediated is a relative measure that is influenced by attenuation of the residual direct association. Concurrently, the direct effect of MVPA attenuated over time, suggesting that depression represents a temporally robust and increasingly dominant pathway linking PA to dementia risk, especially in men. These findings suggest that depression represents a temporally robust pathway linking MVPA to dementia risk, while the remaining direct association likely reflects multiple unmeasured mechanisms. The increasing proportion of the association across longer exclusion windows suggests that psychological pathways may play a key role in the long-term association between MVPA and dementia risk. These findings suggest that the neuroprotective benefits of PA may be partly explained by psychological pathways and partly through cardiometabolic and functional pathways. Our results support strategies that combine PA promotion with targeted depression prevention and early mental health screening, particularly in mid-to-late life, as a potentially effective approach to reducing future dementia risk.

Our study also provides important implications for prevention such as targeting PA and mental health may lead to greater overall benefits. Moreover, combining structured PA programs, such as walking or exercise interventions, with routine screening and management of depression may improve PA adherence and downstream cognitive benefits. Additionally, the contribution of cardiometabolic and functional pathways highlights the importance of comprehensive risk factor management, including control of vascular and metabolic conditions and maintenance of functional capacity. Our results support an integrated prevention framework combining PA promotion, mental health care, and cardiometabolic management to reduce dementia risk.These findings suggest that the benefits of PA may partly depend on its downstream effects on mental health, highlighting depression as a potentially actionable target for dementia prevention. Sex-specific pathways indicate a need for precision intervention, with greater emphasis on cardiometabolic and behavioral risk in males and metabolic–functional pathways in females. This integrated approach is supported by the central role of depression as a mediator, accounting for a meaningful proportion of the association and becoming more prominent over longer follow-up, particularly among males.
Key strengths of this study include device-based objective MVPA, LLM-guided feature selection based causal discovery, and robust confounder adjustment (e.g., age, sex, education) for chain mediation analysis. Our study demonstrates that MVPA influences dementia through behavioral, cardiometabolic, and psychological pathways. 
We integrated large-scale data with causal discovery and chain mediation analyses to identify key pathways and quantify their relative contributions, providing a more mechanistic interpretation than conventional association-based approaches. Our LLM-guided feature selection approach enables scalable, knowledge-informed modeling in high-dimensional epidemiological data that could be generalized to other tasks and domains. Our findings highlight depression as a modifiable intervention target and support prevention strategies combining PA promotion, mental health care, and cardiometabolic risk management with sex-specific focused pathway differences. However, limitations include the observational nature of the data limits definitive causal inference, and residual confounding cannot be excluded. Second, accelerometer data was only collected for seven days which reflects short-term behavior and may not capture lifetime activity patterns. Furthermore, LLM-guided approach and manual curation, may be sensitive to model assumptions and input data quality which could affect reproducibility. Disease classification based on ICD codes may not fully capture disease severity. Despite the large sample size, generalizability may be influenced by cohort-specific characteristics of the UK Biobank, and external validation in more diverse populations is warranted. Although our analyses support depression as a mediator, bidirectional relationships between depression and MVPA cannot be fully excluded in observational data. Nevertheless, our results support multidimensional prevention strategies that combine PA promotion, depression screening and treatment, and cardiometabolic risk management to reduce the burden of dementia in aging populations.


\section{Methods}

\subsection*{Study design and participants}
Data was obtained from the UK Biobank, a prospective population-based cohort comprising extensive phenotypic, lifestyle, and health-related measures. We utilized accelerometer-derived PA data from 103,612 participants who completed a 7-day wrist-worn assessment between 2013 and 2015 \cite{doherty2017large}. MVPA was quantified using a machine learning–based approach (Random Forest) \cite{walmsley2022reallocation}. In the current work, we restricted the sample to older adults aged $\geq 60$ years with valid accelerometer data and complete covariate information. Covariates included demographic, socioeconomic, and clinical variables harmonized for causal modeling and mediation analyses.

\subsection*{Inclusion and exclusion criteria}
Participants were excluded if they (1) had withdrawn consent from the UK Biobank, (2) lacked activity data for any hour in the 24-hour cycle, or (3) exhibited excessive nocturnal activity (defined as $>$10\% of daily MVPA between 01:00 and 04:00) \cite{feng2023associations, qian2021association}. In addition, individuals were excluded for having poor-quality accelerometer data, which included unreliable file size (i.e., abnormally small or large; Field ID: 90002); wear time of less than 72 hours or failure to cover all 24-hour periods over the 7-day span (Field ID: 90015); uncalibrated or poorly calibrated devices (Field ID: 90016); devices reused without proper recalibration (Field ID: 90017); data with recording interruptions (Field ID: 90180); or those with excessive data errors, defined as $>$768 anomalies based on interquartile range thresholds (Field ID: 90182). In addition, we excluded participants who were diagnosed with dementia within one year after wearing the accelerometer, to reduce reverse causation. Specifically, all-cause dementia included Alzheimer’s disease, vascular dementia, and other unspecified neurodegenerative dementias (ICD-10 codes F00–F03, F06, F09, G30, G31) \cite{min2024accelerometer} . Additionally, participants with missing data on any of the covariates used in the models were excluded to ensure complete-case analysis.

\begin{figure}
    \centering
    \includegraphics[width=\linewidth, trim={9cm 1cm 9cm 0}, clip]{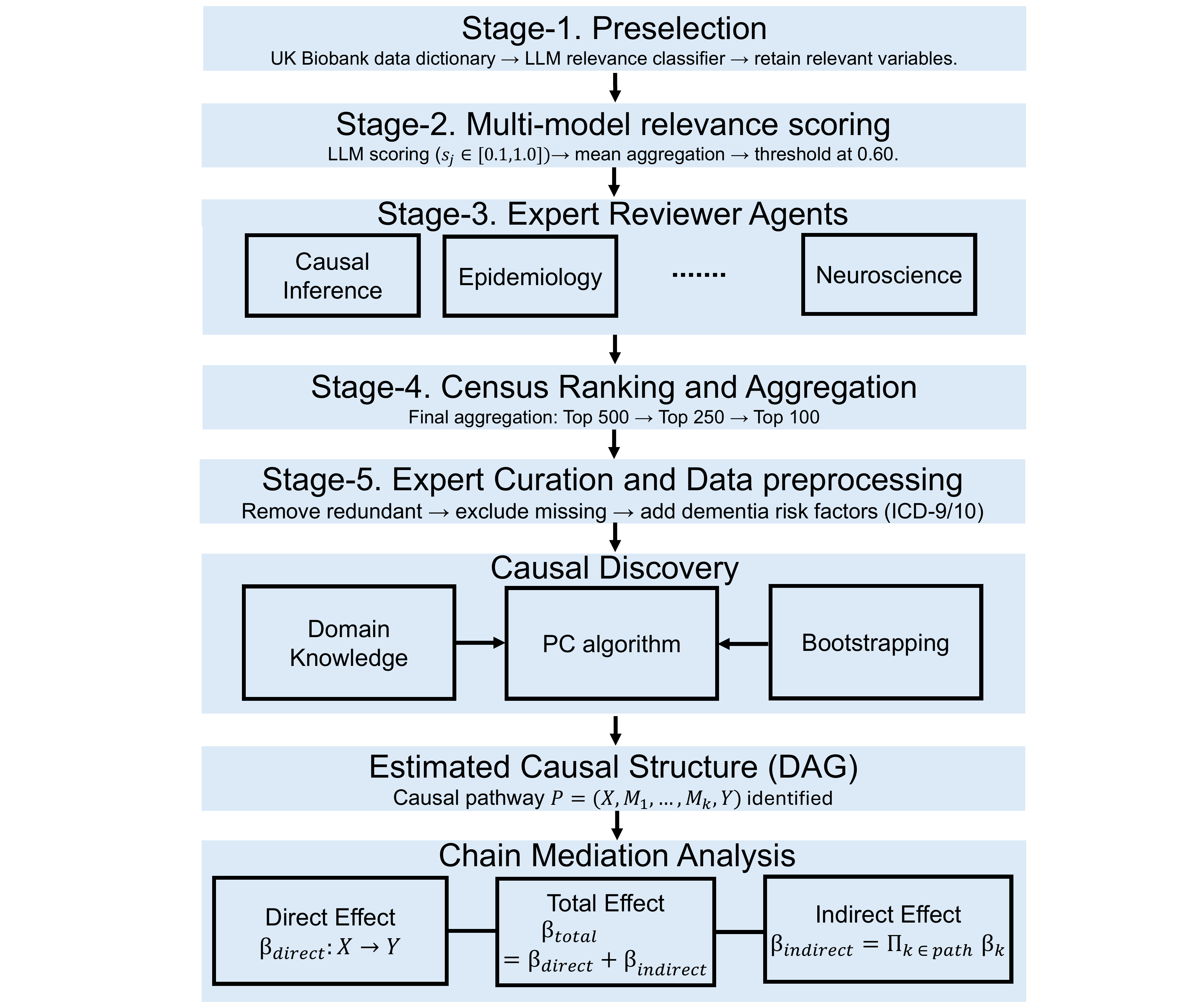}
    \caption{Overview of the variable selection and causal mediation framework. Candidate variables from the UK Biobank were first filtered through a five-stage selection pipeline consisting of LLM-based preselection, multi-model relevance scoring, expert reviewer agents across multiple domains, census ranking and aggregation, and expert curation with data preprocessing. The final curated variables were then used for causal discovery by integrating domain knowledge, the PC algorithm, and bootstrapping to estimate a directed acyclic graph (DAG) representing pathways linking moderate-to-vigorous physical activity (MVPA) to dementia. Identified causal pathways were subsequently evaluated using chain mediation analysis to quantify direct, indirect, and total effects.}
    \label{fig:flowchart}
\end{figure}

\subsection*{LLM-based semantic feature selection}
To systematically identify relevant covariates from the full UK Biobank data dictionary, we implemented a multi-stage large language model (LLM) guided feature selection framework leveraging variable titles and accompanying field descriptions (Fig. \ref{fig:flowchart}). In the first stage (preselection), all candidate variables were evaluated in batches using an LLM-based relevance classifier, which assessed whether each variable was potentially informative for PA–dementia pathways based on its semantic description. Variables classified as relevant were retained for further evaluation. In the second stage (scoring), retained variables were assigned continuous relevance scores $s_j\in [0.1, 1.0]$  using four GPT based models (gpt-4o-mini, gpt-4.1, gpt-oss-120b, and gpt-5), and scores were aggregated. Variables with mean relevance scores below a predefined threshold (0.60) were excluded. To further enhance robustness and domain validity, a reviewer-agent stage was introduced, consisting of multiple expert-informed prompts (e.g., causal inference, epidemiology, neuroimaging, and physical activity science), each generating independent relevance scores. These scores provided a multi-perspective evaluation of each variable’s mechanistic importance (Promopts are provided in supplementary materials). Finally, a consensus-based ranking was performed, and top-ranked variables (e.g., top 500, 250, and 100) were selected using a final aggregation agent for downstream causal modeling. Our hierarchical LLM-driven pipeline enables scalable, knowledge-informed feature selection while reducing subjective bias and improving reproducibility in high-dimensional epidemiological analyses. Following automated ranking, the top 100 variables were evaluated manually to ensure data quality and epidemiological relevance. Variables exhibiting high levels of missingness were excluded to reduce potential bias and instability in downstream causal estimation. Redundant variables were further removed based on pairwise correlation analysis to mitigate multicollinearity and avoid over-representation of highly correlated constructs. Furthermore, established risk factors for dementia—such hearing impairment, and history of brain injury—were included when not already present in the dataset to ensure suitable representation of known clinical determinants. These variables were derived using harmonized ICD-9 and ICD-10 diagnostic codes within the UK Biobank clinical records (Supplementary Table 1).

\subsection*{Causal Discovery and Mediation analysis}
We incorporated domain knowledge,$B$, as structural constraints to ensure that the learned causal structure reflects plausible real-world relationships. For instance, for all$X = (X_1, \ldots, X_p)$ observed variables, we defined exogenous variables $\mathbb{R}$(demographic and socioeconomic factors e.g. Sex, Age, Ethnicity, Qualifications, Townsend deprivation index) such that $\forall X_k \in \mathbb{R}, \; \mathrm{Parent}(X_k) = \emptyset$, and outcome variable (dementia) Y was modeled as a sink node with $\mathrm{Child}(Y) = \emptyset$.  We incorporated PC \cite{spirtes2000causation, du2025causal,sedgewick2019mixed} to estimate causal structure with conditional independence between variables$X_i$ and $X_j$ given a conditioning set $X_i \perp X_j \mid S, \quad S \subseteq X \setminus \{i,j\}$ was tested using Fishers Z\-transform of the partial correlation $\rho_{ij \mid S}$ with $\alpha = 0.05$. To evaluate the robustness of inferred edges, we performed nonparametric bootstrap resampling, re-estimating the graph across multiple resampled datasets with identical settings.
Following causal structure identification, we conducted chain mediation analysis \cite{xu2025linking} using structural equation modeling (SEM) framework to quantify pathway-specific effects . From the estimated causal structure, we defined causal pathway $P = (X, M_1, \ldots, M_k, Y),$ from exposure $X$ , to outcome $Y$ ,  mediated by sequence of mediators $M_i$. Standardized path coefficients $\beta_{ij}$ were estimated using diagonally weighted least squares (DWLS). The indirect effect ($\beta_{\mathrm{indirect}}$)  is obtained as the product of the path-specific effects along the mediation chain while direct effect ($\beta_{\mathrm{direct}}$)  was defined as the standardized coefficient for the edge $X \rightarrow Y$. The total effect ($\beta_{\mathrm{total}}$)  was the sum of direct and indirect effect\cite{baron1986moderator,vanderweele2015explanation,wu2025causal}, $\beta_{\mathrm{total}}=\beta_{\mathrm{direct}}+\beta_{\mathrm{indirect}}.$ The proportion mediated along the specified chain was quantified as \cite{zhang2025frailty}: \[\%\,\mathrm{mediated}=\frac{\beta_{\mathrm{indirect}}}{\beta_{\mathrm{total}}}
\times 100.\]

\subsection*{Covariates}
For mediation analyses, we adjusted for a set of a priori selected covariates based on their established associations with both PA and cognitive outcomes in older adults. These included age at accelerometer, sex, educational attainment (Qualifications), ethnicity, Townsend Deprivation Index, smoking status, and alcohol consumption. Age was modeled as a continuous variable. All covariates were included as adjustment factors in the structural equation modeling framework to reduce potential confounding in the estimated mediation pathways.

\subsection*{Ethics statement}
This study is approved by the ethics of the UK Biobank Ethics Advisory Committee (EAC) and the participants’ right to withdraw samples at any time. The UK Biobank study was performed in accordance with the Declaration of Helsinki. All patients gave written informed consent prior to participation. All methods in this study were conducted with the guidelines and regulations set forth by the UK Biobank and the associated research institutions. Our study utilized data from the UKB accelerometer sub-study, conducted under approved UKB
Application 48388.


\section*{Competing Interests}
The authors declare that the research was conducted in the absence of any commercial or financial relationships that could be construed as a potential conflict of interest.


\section*{Acknowledgements}
The authors are grateful to the UK Biobank for providing access to the dataset used in this study. We also acknowledge the University of Florida for providing computational resources and institutional support. This work is supported by the National Science Foundation under Grant No.2123809; National Institutes of Health/National Institute on Aging (RF1AG071469, R01AG071469); and the National Institutes of Health/National Heart, Lung, and Blood Institute (R01HL159805). The content is solely the responsibility of the authors and does not necessarily represent the official views of the National Institutes of Health and the National Science Foundation.


\section*{Data availability statement}
The data that support the findings of this study are available from the UK Biobank, but restrictions apply to the availability of these data. We used the UK Biobank dataset under the approved application ID  48388. The data is not publicly available but can be accessed from the UK Biobank team upon application approval.

\bibliographystyle{unsrt}
\bibliography{sample}



{\small
\begin{longtable}{%
  >{\raggedright\arraybackslash}p{3.9cm}  
  >{\centering\arraybackslash}p{3.1cm}    
  >{\centering\arraybackslash}p{2.8cm}    
  >{\centering\arraybackslash}p{2.8cm}    
  >{\centering\arraybackslash}p{1.5cm}    
}

\caption{Baseline characteristics of study participants stratified by sex.}
\label{tab:demographics}\\

\toprule
\textbf{Characteristic} &
\textbf{Total} \newline \textbf{(\textit{N} = 42,293)} &
\textbf{Female} \newline \textbf{(\textit{n} = 22,526)} &
\textbf{Male} \newline \textbf{(\textit{n} = 19,767)} &
\textbf{\textit{p}-value} \\
\midrule
\endfirsthead

\multicolumn{5}{l}{\small\textit{Table \ref{tab:demographics} continued}}\\[4pt]
\toprule
\textbf{Characteristic} &
\textbf{Total} \newline \textbf{(\textit{N} = 42,293)} &
\textbf{Female} \newline \textbf{(\textit{n} = 22,526)} &
\textbf{Male} \newline \textbf{(\textit{n} = 19,767)} &
\textbf{\textit{p}-value} \\
\midrule
\endhead

\midrule
\endfoot

\bottomrule
\multicolumn{5}{p{15.9cm}}{\footnotesize Values are presented as mean $\pm$ standard deviation
  for continuous variables and \textit{n} (\%) for categorical and binary variables.
  \textit{p}-values for continuous variables were calculated using the Mann–Whitney U test;
  \textit{p}-values for categorical variables were calculated using the chi-squared test.
  Significance threshold: \textit{p}~$<$~0.05.
 TDI: Townsend
Deprivation Index, CKD:  Chronic Kidney Disease, 
  MVPA, moderate-to-vigorous physical activity; HDL, high-density lipoprotein;
  LDL, low-density lipoprotein; COPD, chronic obstructive pulmonary disease;
  PM10 (particulate matter with diameter less than or equal to 10 micrometres).PM10 (particulate matter with diameter less than or equal to 2.5 micrometres). *Responses of “Prefer not to answer,” “Never,” and “None of the above”
were grouped into a single category.
}\\
\endlastfoot

Age (years)
  & $67.0 \pm 4.1$  & $66.7 \pm 4.1$  & $67.3 \pm 4.2$  & $<$0.001 \\[2pt]
\midrule
TDI
  & $-2.0 \pm 2.7$  & $-1.9 \pm 2.7$  & $-2.0 \pm 2.6$  & $<$0.001 \\[4pt]
\midrule
\textit{Ethnicity}           &               &               &               & 0.677 \\
\quad Others             & 913 (2.2\%)   & 493 (2.2\%)   & 420 (2.1\%)   &       \\
\quad White                  & 41,380 (97.8\%)& 22,033 (97.8\%)& 19,347 (97.9\%)&      \\[4pt]
\midrule
\textit{Qualifications}      &               &               &               & $<$0.001 \\
\quad Never*                  & 4,966 (11.7\%) & 2,640 (11.7\%) & 2,326 (11.8\%) &      \\
\quad A level, O level, or equivalent         & 20,440 (48.3\%)& 11,431 (50.7\%)& 9,009 (45.6\%) &      \\
\quad College/University        & 16,887 (39.9\%)& 8,455 (37.5\%) & 8,432 (42.7\%) &      \\[8pt]
\midrule

Obesity (BMI $\geq$ 30 kg/m$^2$)
  & 8,222 (19.4\%) & 4,086 (18.1\%) & 4,136 (20.9\%) & $<$0.001 \\[2pt]
\midrule
Waist circumference (cm)
  & $89.5 \pm 12.8$ & $83.5 \pm 11.5$ & $96.2 \pm 10.6$ & $<$0.001 \\[2pt]
\midrule
HDL cholesterol (mmol/L)
  & $1.5 \pm 0.4$  & $1.7 \pm 0.4$  & $1.3 \pm 0.3$  & $<$0.001 \\[2pt]
\midrule
LDL cholesterol (mmol/L)
  & $3.6 \pm 0.9$  & $3.8 \pm 0.9$  & $3.5 \pm 0.9$  & $<$0.001 \\[2pt]
\midrule
Triglycerides (mmol/L)
  & $1.7 \pm 0.9$  & $1.6 \pm 0.8$  & $1.9 \pm 1.0$  & $<$0.001 \\[2pt]
\midrule
Systolic blood pressure (mmHg)
  & $142.6 \pm 19.3$ & $140.6 \pm 19.8$ & $144.9 \pm 18.5$ & $<$0.001 \\[2pt]
\midrule
Diastolic blood pressure (mmHg)
  & $82.2 \pm 10.4$ & $80.7 \pm 10.3$ & $83.8 \pm 10.3$ & $<$0.001 \\[8pt]
\midrule

\textit{Smoking status}       &               &               &               & $<$0.001 \\
\quad Never*                   & 22,757 (53.8\%)& 13,326 (59.2\%)& 9,431 (47.7\%) &      \\
\quad Previous                & 17,178 (40.6\%)& 8,215 (36.5\%) & 8,963 (45.3\%) &      \\
\quad Current                & 2,358 (5.6\%) & 985 (4.4\%)   & 1,373 (6.9\%) &       \\[4pt]
\midrule
\textit{Alcohol intake frequency} &           &               &               & $<$0.001 \\
\quad Never*                    & 2,371 (5.6\%) & 1,546 (6.9\%) & 825 (4.2\%)   &       \\
\quad Daily  & 11,229 (26.6\%)& 4,911 (21.8\%)& 6,318 (32.0\%) &      \\
\quad 1–4 times/week       & 20,801 (49.2\%)& 10,665 (47.3\%)& 10,136 (51.3\%)&      \\
\quad Occasionally  & 7,892 (18.7\%)& 5,404 (24.0\%)& 2,488 (12.6\%) &      \\[4pt]
\midrule
Sleep duration (hrs/night)
  & $8.92\pm 1.24$  & $8.92\pm 1.17$  & $8.91 \pm 1.33$  & 0.004 \\[8pt]
\midrule

\textit{Usual walking pace}   &               &               &               & $<$0.001 \\
\quad Slow pace                    & 2,013 (4.8\%) & 1,149 (5.1\%) & 864 (4.4\%)   &       \\
\quad Steady average pace        & 21,635 (51.2\%)& 11,679 (51.8\%)& 9,956 (50.4\%) &      \\
\quad Brisk pace                 & 18,645 (44.1\%)& 9,698 (43.1\%)& 8,947 (45.3\%) &      \\[4pt]
\midrule
\textit{Walking days $\geq$10 min/week} &     &               &               & $<$0.001 \\
\quad 1 day                   & 1,115 (2.6\%) & 489 (2.2\%)   & 626 (3.2\%)   &       \\
\quad 2 days                  & 2,550 (6.0\%) & 1,192 (5.3\%) & 1,358 (6.9\%) &       \\
\quad 3 days                  & 3,613 (8.5\%) & 1,917 (8.5\%) & 1,696 (8.6\%) &       \\
\quad 4 days                  & 3,890 (9.2\%) & 2,095 (9.3\%) & 1,795 (9.1\%) &       \\
\quad 5 days                  & 6,821 (16.1\%)& 3,654 (16.2\%)& 3,167 (16.0\%) &      \\
\quad 6 days                  & 4,692 (11.1\%)& 2,492 (11.1\%)& 2,200 (11.1\%) &      \\
\quad 7 days                  & 19,612 (46.4\%)& 10,687 (47.4\%)& 8,925 (45.2\%)&      \\[4pt]
\midrule
Duration of walks (min)
  & $50.7 \pm 64.8$ & $49.0 \pm 62.8$ & $52.7 \pm 67.0$ & $<$0.001 \\[2pt]
\midrule
Average sedentary time (Hours/day) 
  & $9.36 \pm 1.76$ & $9.06 \pm 1.68$ & $9.71 \pm 1.79$ & $<$0.001 \\[2pt]
\midrule
Average light activity (min/day)
  & $303.1 \pm 95.1$ & $327.7 \pm 91.9$ & $275.0 \pm 90.7$ & $<$0.001 \\[2pt]
\midrule
Average MVPA (min/day)
  & $40.2 \pm 34.0$ & $33.3 \pm 28.7$ & $48.0 \pm 37.7$ & $<$0.001 \\[2pt]
\midrule
Time spent driving (Hours/day)
  & $0.6 \pm 1.0$  & $0.4 \pm 0.8$  & $0.8 \pm 1.2$  & $<$0.001 \\[8pt]
\midrule

Reaction time (ms)
  & $563.9 \pm 108.5$ & $574.2 \pm 109.9$ & $552.2 \pm 105.6$ & $<$0.001 \\[4pt]
\midrule
\textit{Leisure/social activities} & &               &               & $<$0.001 \\
\quad None                       & 11,260 (26.6\%)& 5,658 (25.1\%)& 5,602 (28.3\%) &      \\
\quad Religious group                       & 2,447 (5.8\%) & 1,403 (6.2\%) & 1,044 (5.3\%) &       \\
\quad Adult education class                       & 987 (2.3\%)   & 713 (3.2\%)   & 274 (1.4\%)   &       \\
\quad Pub or social club                       & 4,084 (9.7\%) & 1,238 (5.5\%) & 2,846 (14.4\%)&       \\
\quad Sports club or gym                       & 5,664 (13.4\%)& 3,031 (13.5\%)& 2,633 (13.3\%) &      \\
\quad Multiple activities               & 17,851 (42.2\%)& 10,483 (46.5\%)& 7,368 (37.3\%)&      \\[8pt]
\midrule

Depression
  & 1,639 (3.9\%)  & 1,029 (4.6\%)  & 610 (3.1\%)   & $<$0.001 \\[2pt]
\midrule
Heart disease
  & 6,092 (14.4\%) & 2,292 (10.2\%) & 3,800 (19.2\%) & $<$0.001 \\[2pt]
\midrule
Hypertension
  & 11,843 (28.0\%)& 5,437 (24.1\%) & 6,406 (32.4\%) & $<$0.001 \\[2pt]
\midrule
Stroke
  & 815 (1.9\%)   & 289 (1.3\%)   & 526 (2.7\%)   & $<$0.001 \\[2pt]
\midrule
Diabetes
  & 2,576 (6.1\%) & 970 (4.3\%)   & 1,606 (8.1\%) & $<$0.001 \\[2pt]
\midrule
Asthma
  & 4,707 (11.1\%)& 2,629 (11.7\%)& 2,078 (10.5\%)& $<$0.001 \\[2pt]
\midrule
COPD
  & 939 (2.2\%)   & 429 (1.9\%)   & 510 (2.6\%)   & $<$0.001 \\[2pt]
\midrule
Chronic kidney disease
  & 2,459 (5.8\%) & 1,047 (4.6\%) & 1,412 (7.1\%) & $<$0.001 \\[2pt]
\midrule
Cancer
  & 7,944 (18.8\%)& 4,326 (19.2\%)& 3,618 (18.3\%)& 0.019 \\[2pt]
\midrule
Arthritis
  & 8,080 (19.1\%)& 4,647 (20.6\%)& 3,433 (17.4\%)& $<$0.001 \\[2pt]
\midrule
Hearing impairment
  & 1,460 (3.5\%) & 638 (2.8\%)   & 822 (4.2\%)   & $<$0.001 \\[2pt]
\midrule
Brain injury
  & 1,269 (3.0\%) & 612 (2.7\%)   & 657 (3.3\%)   & $<$0.001 \\[8pt]
\midrule

PM$_{10}$ (2010, $\mu$g/m$^3$)
  & $16.1 \pm 1.9$ & $16.1 \pm 1.9$ & $16.1 \pm 1.9$ & 0.006 \\[2pt]
\midrule
PM$_{2.5}$ (2010, $\mu$g/m$^3$)
  & $9.8 \pm 1.0$  & $9.8 \pm 1.0$  & $9.8 \pm 1.0$  & 0.004 \\[8pt]
\midrule

Incident dementia
  & 647 (1.5\%)   & 277 (1.2\%)   & 370 (1.9\%)   & $<$0.001 \\
\end{longtable}
}

\clearpage


\section*{Supplementary material}

# **Supplementary Materials**: Causal discovery identifies pathways linking physical activity to dementia risk in the UK Biobank

## Contents



*Table S. 1. Names and corresponding UK Biobank field IDs/ICD codes for the variables used in this study.*

| Feature Name | UK Biobank Field ID |
|---|---|
| Sex | 31 |
| Townsend Deprivation Index | 22189 |
| Sleep Duration | 1160 |
| HDL Cholesterol | 23406 |
| LDL Direct | 30780 |
| Diastolic Blood Pressure (DBP) | 4079 |
| Systolic Blood Pressure (SBP) | 4080 |
| Waist Circumference | 48 |
| Triglycerides | 30870 |
| PM10 (2010) | 24006 |
| PM2.5 (2010) | 24005 |
| Reaction Time | 20023 |
| Year of birth | 34 |
| Month of birth | 52 |
| Start time of wear | 90010 |
| Obesity (BMI source) | 21001 |
| Ethnic Background | 21000 |
| Smoking Status | 20116 |
| Alcohol Intake Frequency | 1558 |
| Qualifications | 6138 |
| Avg Sedentary Per Day (Mins) | 40043 |
| Avg MVPA Per Day (Mins) | 40044 |
| Avg Light Per Day (Mins) | 40045 |
| Usual Walking Pace | 924 |
| Number days/week walked 10+ min | 864 |
| Duration of Walks | 874 |
| Time Spent Driving | 1090 |
| Leisure/Social Activities | 6160 |

| | |
|---|---|
| Heart Disease* | **ICD-10:** I20, I21, I22, I23, I24, I25, I42, I50, I46, I49, I51, I52, I67, I68, I69, I70<br><br>**ICD-9:** 413, 410, 411, 412, 414, 425, 428, 393, 394, 395, 396, 397, 398, 399, 400, 420, 421, 422, 423, 424, 426, 427, 429 |
| Stroke* | **ICD-10:** I64, I65, I60, I61, I62, I66, I63<br><br>**ICD-9:** 438,435,430,431,432,433,434,436,437 |
| Diabetes* | **ICD-10:** G632, G590, H360, H280, E10, E11, E12, E13, E14, E00<br><br>**ICD-9:** 250, 240 |
| Asthma* | **ICD-10:** J45, J450, J451, J458, J459<br><br>**ICD-9:** 493 |
| COPD* | **ICD-10:** J41, J42, J43, J44<br><br>**ICD-9:** 491,492,494 |
| Chronic Kidney Disease* | **ICD-10:** Q611, Q612, Q613, N17, N18, N19, E112, N028<br><br>**ICD-9:** 75315,584,585,586,587,588 |
| Cancer* | **ICD-10:** C00, C01, C02, C03, C04, C05, C06, C07, C08, C09, C10, C11, C12, C13, C14, C15, C16, C17, C18, C19, C20, C21, C22, C23, C24, C25, C26, C30, C31, C32, C33, C37, C38, C39, C40, C41, C42, C43, C44, C45, C46 ,C47, C48, C49, C50, C51, C52, C53, C54, C55, C57, C58, C60, C61, C62, C63, C64, C65, C66, C67, C68, C69, C70, C71, C72, C73, C74, C75, C76, C77, C78, C79, C80, C81, C82, C83, C84, C85, C86, C88, C90, C91, C92, C93, C94, C95, C96, C97<br>**ICD-9:** 162, 173,172,151,150,153,154,185,183, 174,182, 140,141, 142,143,144, 145,146,147,148,149, 152, 155, 156, 157, 158, 159, 160, 161, 163, 164, 165, 166, 167, 171, 175, 176,177, 180, 181, 184, 186, 187, 188, 189, 190, 191, 192, 193, 194, 195, 196, 197, 198, 200, 201, 202, 203, 204, 205, 206, 207, 208 |
| Arthritis* | **ICD-10:** M4910, M4920, M053, M058, M059, M06, M15, M16, M17, M18, M19 |
| Brain Injury* | **ICD-10:** S00, S01, S02, S03, S04, S05, S06, S07, S08, S09, S020, S021, S028, S0291, S0402, S0403, S0404, S071, T744<br>**ICD-9:** 95901, 800, 801, 802, 803, 804 |
| Hearing Impairment* | **ICD-10:** H90, H91<br><br>**ICD-9:** 389 |
| Hypertension* | **ICD-10:** I10, I11, I110, I119, I12, I120, I129, I13, I130, I131, I132, I139, I15, I150, I151, I152, I158, I159<br>**ICD-9:** 401,402,403,404,405 |
| Depression* | **ICD-10:** F32, F33, F341, F381, F204 |

| | |
|---|---|
| | **ICD-9:** 2962, 2963, 3004, 311 |
| Dementia | 130836, 130838, 130840, 130842, 130848, 131036, 131038 |

* Disease status was defined using ICD-9 and ICD-10 diagnostic codes. Participants were classified as having the condition if the first recorded diagnosis occurred on or before the start of wearable accelerometer (baseline). Participants without a diagnosis prior to baseline were classified as not having the condition.

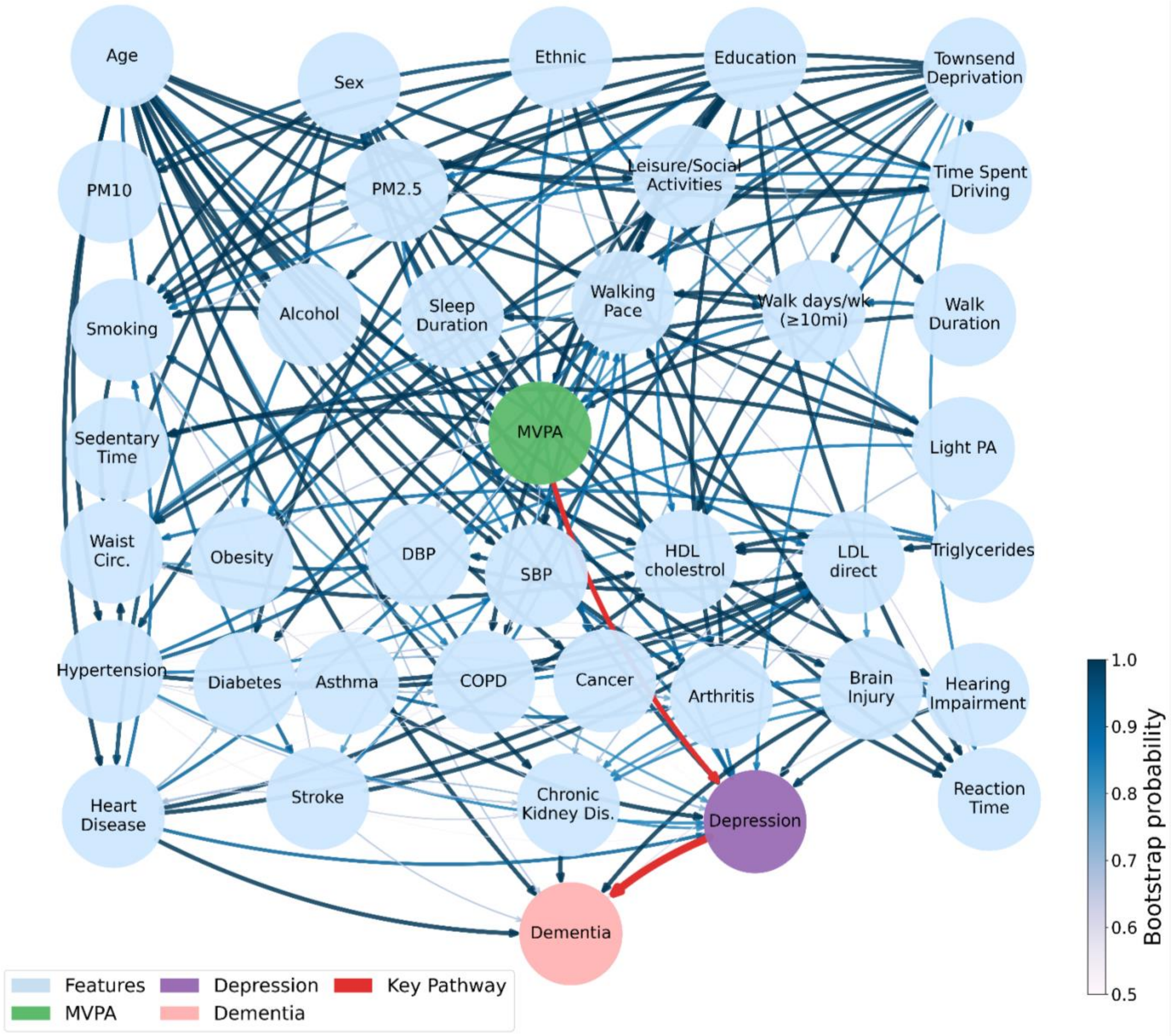


*Figure S. 1. Partially directed acyclic graph (PDAG) estimated using the PC algorithm after excluding participants who developed dementia within 1 year of accelerometer/device wear. Nodes represent study variables, and edges represent conditional dependency relationships remaining after iterative conditional independence testing. Edge color and width indicate bootstrap probability, with darker and thicker edges denoting more stable relationships across bootstrap samples. MVPA (green), depression (purple), and dementia (pink) are highlighted, and the red pathway denotes the principal pathway of interest identified in the network. Edge directions reflect orientations inferred by the PC algorithm.*

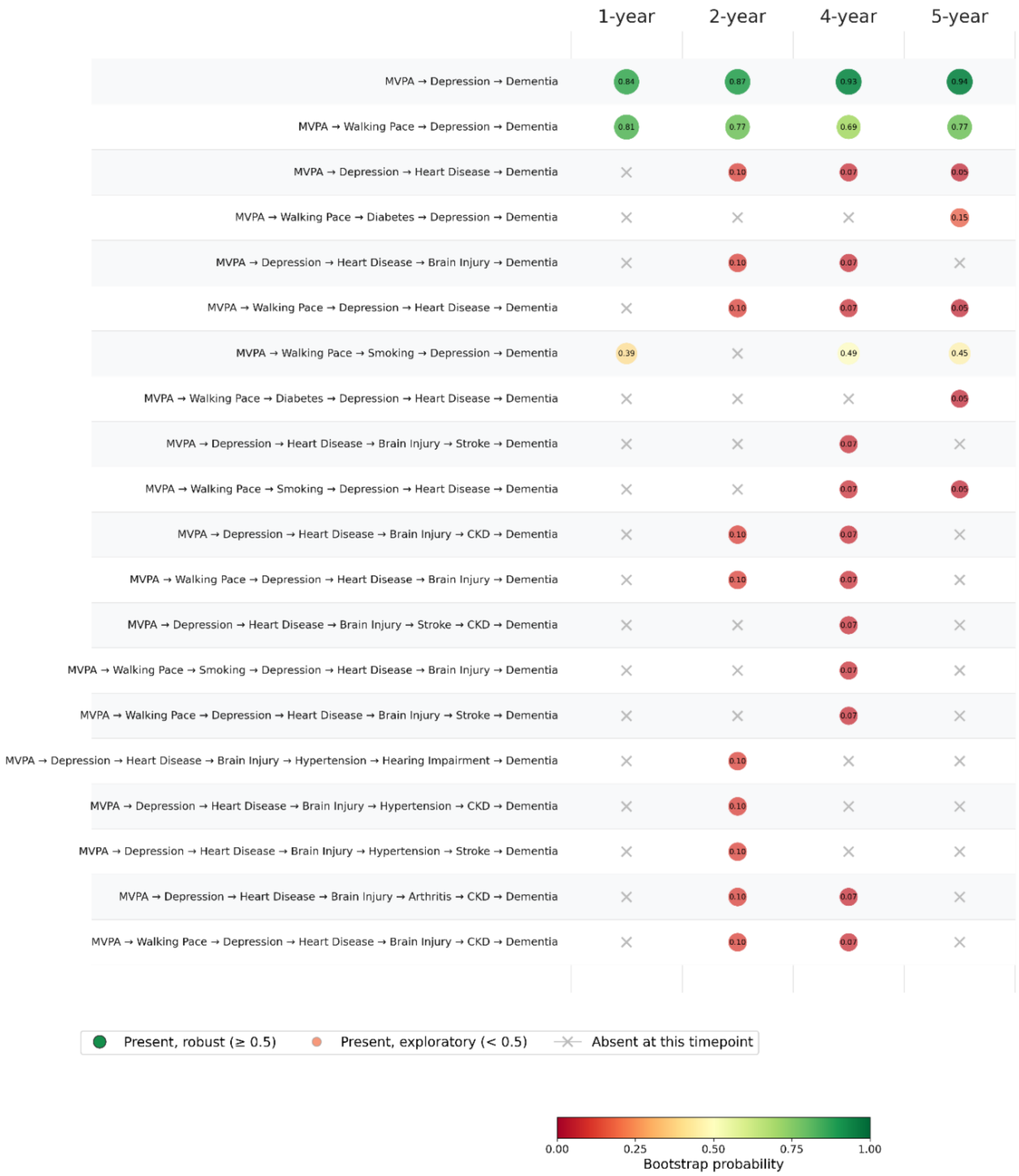


*Figure S. 2. Dot plot showing unique causal pathways from moderate-to-vigorous physical activity (MVPA) to dementia identified by the PC algorithm, across 1-, 2-, 4-, and 5-year exclusion windows. Circle size and color indicate the minimum bootstrap selection probability among all edges in the pathway, ranging from red (low stability) to green (high stability), based on 200 bootstrap resamples. Values shown inside circles are the minimum bootstrap probabilities. A cross (×) indicates the pathway was not identified at that timepoint.*

*Table S. 2.* ***Mediation analysis of the depression pathway.*** *Direct, indirect, and total effects of moderate-to-vigorous physical activity (MVPA) on dementia risk through depression, stratified by sex and exclusion periods (1, 2, and 4 years). The table presents the direct effect ($\beta_{direct}$), indirect effect mediated by depression ($\beta_{indirect}$), total effect ($\beta_{total}$), and percentage mediated for the overall population, males, and females.*

| Population | Exclusion | Direct effect | Indirect effect | | | Total effect $\beta_{\text{total}}$ | % Mediated |
|---|---|---|---|---|---|---|---|
| | | MVPA → Dementia $\beta_{\text{direct}}$ | MVPA → Depression | Depression → Dementia | Total indirect $\beta_{\text{indirect}}$ | | |
| Overall | 1 years | -0.0155 | -0.052 | 0.053 | -0.0028 | -0.0183 | 15.06% |
| | 2 years | -0.0146 | -0.052 | 0.053 | -0.0028 | -0.0174 | 16.06% |
| | 4 years | -0.0106 | -0.053 | 0.047 | -0.0025 | -0.0130 | 19.08% |
| | 5 years | -0.009 | -0.053 | 0.051 | 0.0027 | -0.0113 | 23.77% |
| Male | 1 years | -0.0139 | -0.051 | 0.066 | -0.0033 | -0.0172 | 19.31% |
| | 2 years | -0.0118 | -0.051 | 0.068 | -0.0035 | -0.0153 | 22.58% |
| | 4 years | -0.0054 | -0.052 | 0.056 | -0.0029 | -0.0083 | 35.03% |
| | 5 years | -0.0019 | -0.052 | 0.056 | -0.0029 | -0.0048 | 60.75% |
| Female | 1 years | -0.0167 | -0.054 | 0.044 | -0.0023 | -0.0191 | 12.27% |
| | 2 years | -0.0171 | -0.054 | 0.042 | -0.0023 | -0.0194 | 11.82% |
| | 4 years | -0.0176 | -0.055 | 0.042 | -0.0023 | -0.0199 | 11.59% |
| | 5 years | -0.0177 | -0.055 | 0.048 | -0.0026 | -0.0203 | 13.00% |

*Table S. 3.* ***Chain mediation analysis of the walking pace–depression pathway.*** *Direct, indirect, and total effects of moderate-to-vigorous physical activity (MVPA) on dementia risk through the sequential mediators of walking pace and depression (MVPA → walking pace → depression → dementia), stratified by sex and exclusion periods (1, 2, and 4 years). The table presents the pathway-specific indirect effects, direct effect ($\beta_{direct}$), total indirect effect ($\beta_{indirect}$), total effect ($\beta_{total}$), and percentage mediated for the overall population, males, and females.*

| Population | Exclusion | Direct effect | Indirect effect | | | | Total effect $\beta_{\text{total}}$ | % Mediated |
|---|---|---|---|---|---|---|---|---|
| | | MVPA → Dementia $\beta_{\text{direct}}$ | MVPA → Walking pace | Walking pace → Depression | Depression → Dementia | Total indirect $\beta_{\text{indirect}}$ | | |
| Overall | 1 years | -0.0164 | 0.176 | -0.056 | 0.054 | -0.0005 | -0.0169 | 3.13% |
| | 2 years | -0.0151 | 0.176 | -0.055 | 0.054 | -0.0005 | -0.0156 | 3.37% |
| | 4 years | -0.0112 | 0.176 | -0.055 | 0.049 | -0.0005 | -0.0117 | 4.04% |
| | 5 years | -0.010 | 0.176 | -0.055 | 0.052 | -0.0005 | -0.0102 | 4.99% |
| Male | 1 years | -0.0146 | 0.135 | -0.047 | 0.067 | -0.0004 | -0.0150 | 2.83% |
| | 2 years | -0.0125 | 0.134 | -0.047 | 0.069 | -0.0004 | -0.0129 | 3.41% |
| | 4 years | -0.0061 | 0.134 | -0.047 | 0.057 | -0.0004 | -0.0065 | 5.49% |
| | 5 years | -0.003 | 0.134 | -0.047 | 0.058 | -0.0004 | -0.0035 | 10.56% |
| Female | 1 years | -0.0172 | 0.219 | -0.061 | 0.044 | -0.0006 | -0.0178 | 3.34% |
| | 2 years | -0.0171 | 0.219 | -0.061 | 0.042 | -0.0006 | -0.0177 | 3.16% |
| | 4 years | -0.0172 | 0.219 | -0.060 | 0.043 | -0.0006 | -0.0177 | 3.17% |
| | 5 years | -0.0175 | 0.219 | -0.060 | 0.048 | -0.0006 | -0.0182 | 3.48% |

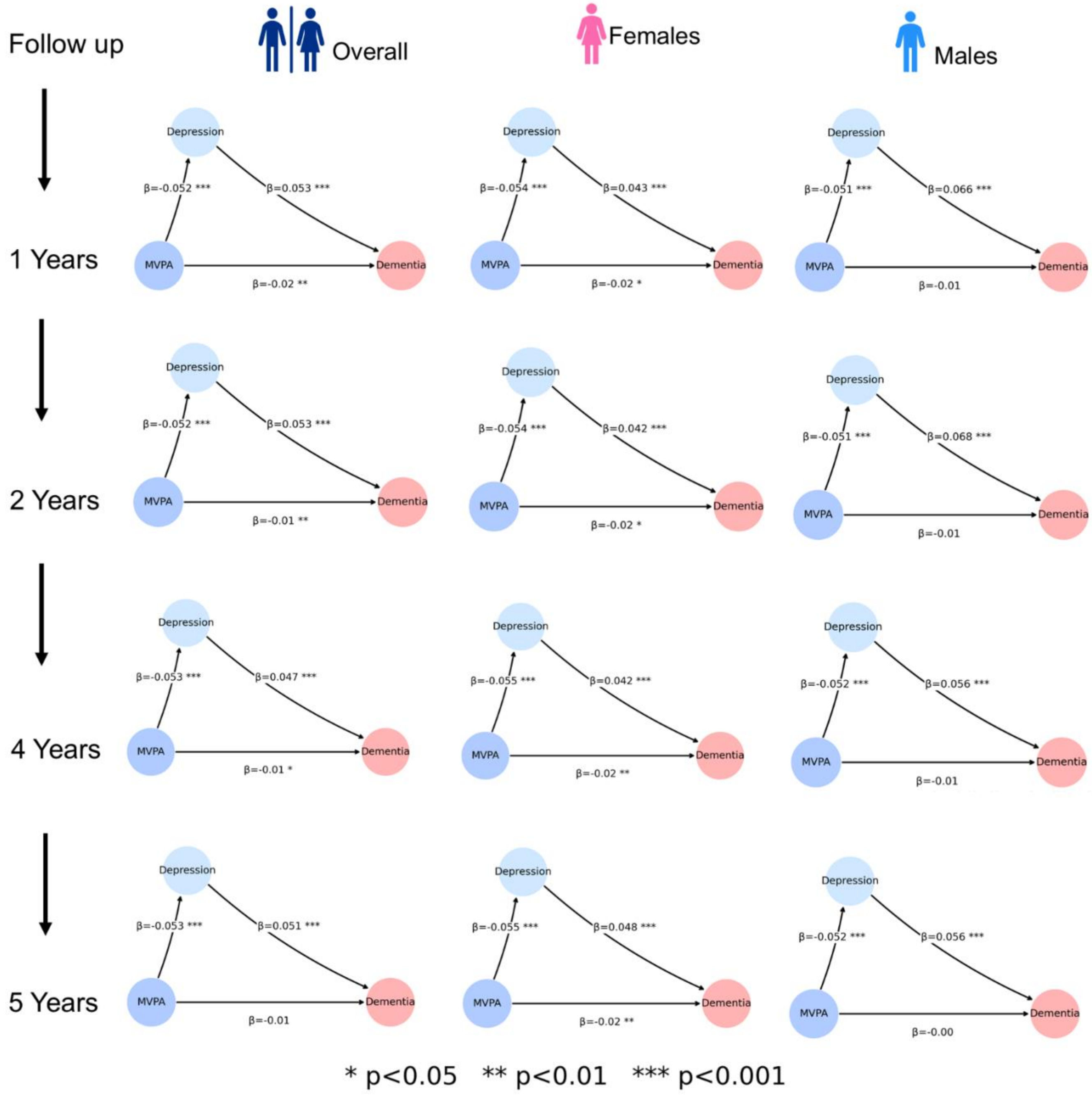


*Figure S. 3. Depression-mediated pathways linking moderate-to-vigorous physical activity (MVPA) to dementia risk across progressively longer follow-up exclusion windows. Structural equation mediation models evaluating the pathway* ***MVPA → Depression → Dementia*** *in the overall cohort (left), females (middle), and males (right) after excluding incident dementia cases occurring within 1, 2, 4, and 5 years following accelerometer assessment. The magnitude of the direct MVPA–dementia association attenuated with longer exclusion periods, whereas the relative contribution of the depression-mediated pathway increased, particularly among males.*

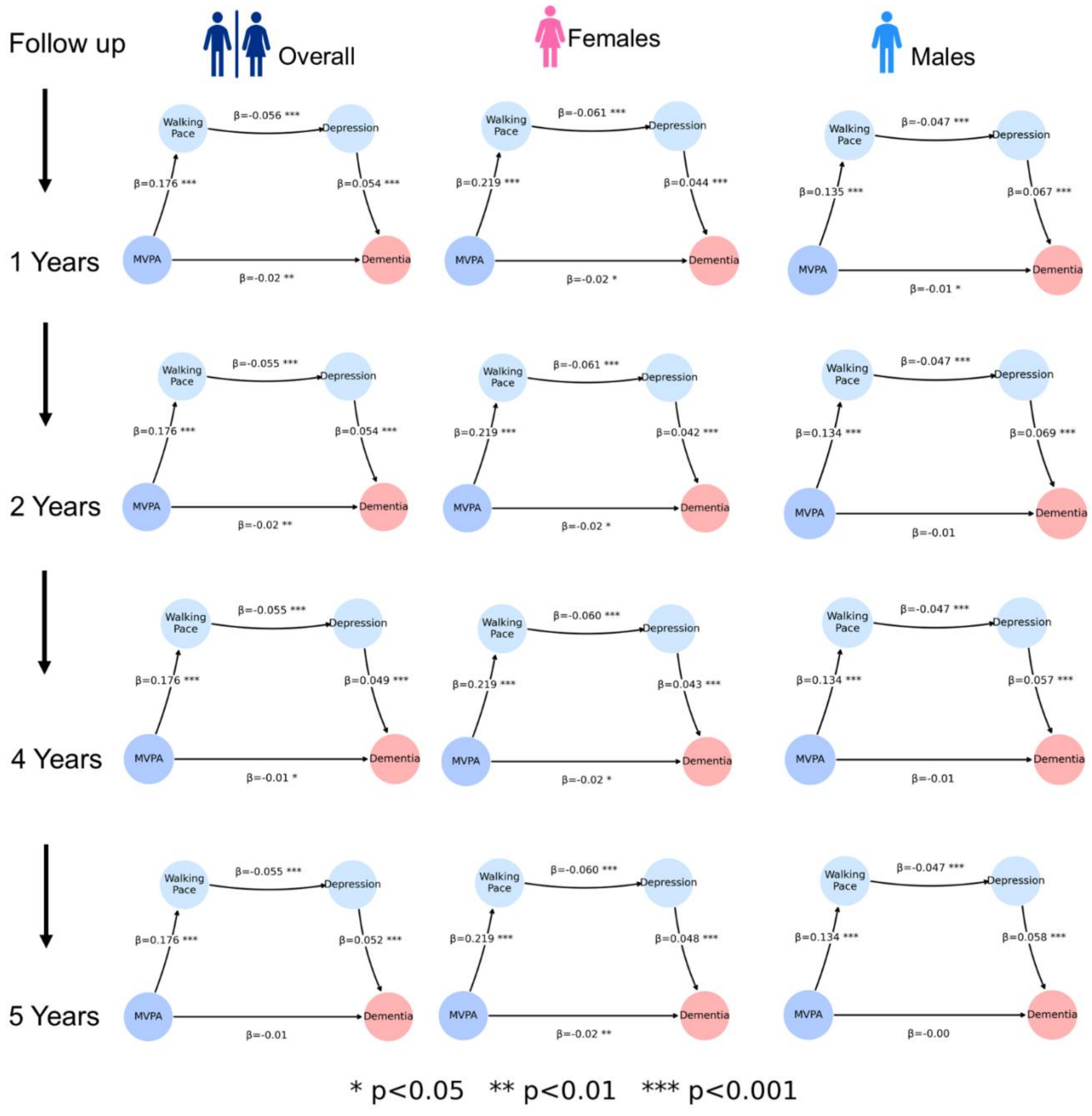


*Figure S. 4. Structural equation mediation models evaluating the pathway* ***MVPA →Walking Pace → Depression → Dementia*** *in the overall cohort (left), females (middle), and males (right) after excluding incident dementia cases occurring within 1, 2, 4, and 5 years following accelerometer assessment.*

## Prompts

**Preselect Variables:**

You are an expert in aging, cognition, and physical activity, with a focus on the impact of physical activity on dementia.

Task: For each UK Biobank variable below, decide if it is directly and clearly relevant to studying the impact of physical activity on dementia, or if it has a causal pathway linking physical activity to dementia risk, progression, or brain aging.

Consider:

- variable name

- variable description (notes)

- known biomedical pathways

- clinical literature on physical activity and dementia

Output format — ONLY one of the following for each variable (no justification, no extra text):

Yes (directly relevant or involved in a causal pathway)

No (not relevant).

**Pre-select Score:**

You are an expert in aging, cognition, and physical activity, with a focus on dementia research. Thik step by step for below:

Task: For each UK Biobank variable below, assign a numeric relevance score between 0.1 (very unlikely relevant) and 1.0 (highly relevant) indicating how important this variable is for studying causal relationships between physical activity and dementia/brain aging.

Output format:

- Only output the variable name followed by its score, separated by a colon, one per line.

- Example:

Sex: 1.0

Var456: 0.3.

**Reviewer agents:**

1. You are a causal-inference expert in aging and dementia research. Think step-by-step.

Task: For each variable below, assign a numeric relevance score between 0.10 (very unlikely relevant) and 1.00 (highly relevant) for inclusion in a causal model of:  physical activity (exposure), cognition/brain aging (mediators/outcomes), dementia (endpoint).

Consider: whether the variable is an exposure measure (PA), outcome (cognitive test, MRI), essential confounder, or plausible mediator. Base judgments on established scientific evidence and causal reasoning.

Output rules: ONLY lines of the form `VariableName: 0.00` (two-decimal score). No extra text, numbering, or justification.

2. You are a neuroscientist specializing in brain aging and MRI biomarkers. Think step-by-step.

Task: Score each variable (0.10–1.00) for its relevance to the pathway: physical activity, brain structure/function (MRI, DTI, functional measures), cognition, and dementia.

Prioritize imaging measures, cognitive outcomes, and critical demographic/genetic confounders. Use neuroscience literature and mechanistic plausibility in your reasoning.

Output rules: ONLY `VariableName: 0.00` lines (two-decimal score), one per line, no commentary.

3. You are a dementia epidemiologist. Think step-by-step.

Task: For each variable, provide a relevance score (0.10–1.00) for modeling how physical activity affects cognitive decline and dementia risk.

Consider established risk factors (age, education, APOE), vascular and metabolic mediators, measurement quality (objective PA vs self-report), and temporality for causal inference.

Output rules: ONLY `VariableName: 0.00` lines (two-decimal score), one per line, nothing else.

4. You are an expert in physical activity measurement and movement science. Think step-by-step.

Task: Score (0.10–1.00) each variable for its utility in studying causal links between physical activity (especially accelerometer-derived metrics, gait, fitness) and brain health/cognition.

Prioritize objective PA measures, mobility metrics, and variables that help adjust for confounding of activity–cognition associations.

Output: ONLY `VariableName: 0.00` lines (two-decimal), one per line.

5. You are a neuroimaging data scientist experienced with UK Biobank MRI. Think step-by-step.

Task: Assign a relevance score (0.10–1.00) to each variable for modeling of physical activity, MRI biomarkers (regional volumes, cortical thickness, white matter integrity), cognition, and dementia.

Favor well-validated imaging phenotypes and essential covariates required for interpretable imaging associations.

Output: ONLY `VariableName: 0.00` (two decimals), one per line.

6. You are a cardiovascular and metabolic aging researcher. Think step-by-step.

Task: Score each variable (0.10–1.00) for its role in vascular/metabolic pathways linking lifestyle factors to brain aging and cognition.

Output: ONLY `VariableName: 0.00` lines (two-decimal), one per line.

7. You are a machine-learning researcher specializing in causal feature selection. Think step-by-step.

Task: For each variable, assign a relevance score (0.10–1.00) that reflects its causal relevance (not merely predictive correlation) for:  physical activity, cognition, brain aging, and dementia.

Penalize redundant measures; favor variables that improve identifiability of causal effects (clear temporality, measurement fidelity, domain knowledge).

Output: ONLY `VariableName: 0.00` (two-decimal) lines, one per line.

8. You are a mediation and mechanistic pathways expert. Think step-by-step.

Task: Score variables (0.10–1.00) by how likely they are to be meaningful mediators, confounders, exposures, or outcomes in the pathway of lifestyle factors with dementia. Such as physical activity, biological mediators (inflammation, sleep, vascular health, neurotrophic factors), brain aging, cognition, and dementia.

Base choices on mechanistic plausibility from existing literature.

Output: ONLY `VariableName: 0.00` lines (two decimals), one per line.

9. You are a cognitive neuroscientist focused on memory and executive function. Think step-by-step.

Task: Assign a relevance score (0.10–1.00) to each variable for modeling of physical activity, cognitive domains (memory, executive function, processing speed), and dementia progression.

Prioritize robust cognitive measures, hippocampal/ frontal imaging markers, and core demographics/genetics.

Output: ONLY `VariableName: 0.00` lines, two decimal scores, one per line.

10. You are a public health and preventive-aging expert. Think step-by-step.

Task: Score variables (0.10–1.00) for their public-health relevance in causal models linking physical activity to cognitive aging and dementia. Emphasize modifiable exposures, scalable measures, and confounders important for policy-relevant inference.

Output: ONLY `VariableName: 0.00` lines (two decimals), one per line, no extra text.

Final agent
You are an expert in aging, cognition, dementia, and physical activity research.

You are given a list of UK Biobank variables, each with descriptive notes.

Your task:

- Carefully rank and order the variables according to their relevance for studying causal relationships between physical activity, cognition, and brain aging.

- Base your ranking on established scientific evidence and reasoning from aging neuroscience, cognitive epidemiology, and precision brain health research.

- Output an ordered list of {target_count} variable names, one per line, without numbering or extra text.



\end{document}